\documentclass[prr,superscriptaddress,reprint,showpacs,longbibliography]{revtex4-2}[prr]
\usepackage{color}
\usepackage{graphicx,textcomp,amssymb,amsmath,dcolumn,hyperref}
\usepackage{siunitx}
\usepackage{comment}
\usepackage{physics}
\usepackage{ulem}

\usepackage{romannum}

\begin{document}

\title{Counting Statistics of Single Electron Transport in Bilayer Graphene Quantum Dots}

\author{Rebekka Garreis}
\email{garreisr@phys.ethz.ch}
\author{Jonas Daniel Gerber}
\affiliation{Solid State Physics Laboratory, ETH Zurich, 8093 Zurich, Switzerland}
\author{Veronika Stará}
\affiliation{Central European Institute of Technology, Brno University of Technology, 612 00 Brno, Czech Republic}
\author{Chuyao Tong}
\affiliation{Solid State Physics Laboratory, ETH Zurich, 8093 Zurich, Switzerland}
\author{Carolin Gold}
\affiliation{Solid State Physics Laboratory, ETH Zurich, 8093 Zurich, Switzerland}
\affiliation{Department of Physics and Astronomy, Columbia University, New York, NY, USA}
\author{Marc Röösli}
\affiliation{Solid State Physics Laboratory, ETH Zurich, 8093 Zurich, Switzerland}

\author{Kenji Watanabe}
\author{Takashi Taniguchi}
\affiliation{National Institute for Material Science, 1-1 Namiki, Tsukuba 305-0044, Japan}
\author{Klaus Ensslin}
\author{Thomas Ihn}
\affiliation{Solid State Physics Laboratory, ETH Zurich, 8093 Zurich, Switzerland}
\author{Annika Kurzmann}
\affiliation{Solid State Physics Laboratory, ETH Zurich, 8093 Zurich, Switzerland}
\affiliation{2nd Institute of Physics, RWTH Aachen University, Aachen, 52074, Germany}

\date{\today}

\begin{abstract}
%The full counting statistics give direct access to the distribution function of current fluctuations. 
%Ist dieser erste Satz notwendig?

We measure telegraph noise of current fluctuations in an electrostatically defined quantum dot in bilayer graphene by real-time detection of single electron tunneling with a capacitively coupled neighboring quantum dot. Suppression of the second and third cumulant (related to shot noise) in a tunable graphene quantum dot is demonstrated experimentally. With this method we demonstrate the ability to measure very low current and noise levels. Furthermore, we use this method to investigate the first spin excited state, an essential prerequisite to measure spin relaxation.

\end{abstract}

\maketitle

\section{Introduction}

The physics of low-dimensional quantum systems has been studied in various semiconductors, such as GaAs \cite{Ciorga2000, Hofmann2016}, InAs~\cite{Shorubalko2006,Fuhrer2007}, and silicon \cite{Wang1994,Culcer2009,Leon2020}. Graphene provides a new and interesting platform, because of its natural two-dimensional character, its specific band structure and the additional valley degeneracy \cite{McCann2007bandstructure,Xiao2010,Varlet2015,Knothe2018}. With rapidly advancing developments in the fabrication process of clean and high quality samples~\cite{Overweg2018, Yue2015}, electrostatically defined quantum dots have been demonstrated~\cite{Eich2018, Banszerus2018, Eich2018doubledot, Banszerus2020singleelectron, Eich2020,  Garreis2021, Banszerus2022}. These works include the investigation of the excited state spectrum~\cite{Kurzmann219excited}, the Kondo effect~\cite{Kurzmann2021}, the Pauli spin and valley blockade~\cite{Tong2022}, as well as the \textit{in situ} tuning from an electron-like to a hole-like quantum dot at a constant bulk density~\cite{Banszerus2020, Tong2021}. Furthermore, capacitively coupling the investigated quantum dot to a neighboring dot allows for charge detection~\cite{Kurzmann2019Chargdet}. Most recently the spin relaxation time $T_1$ in such a system was measured using the Elzerman readout technique and performing single-shot measurements for spin to charge conversion~\cite{GachterGarreis2022}. Time-resolved charge detection~\cite{Guttinger2011} is essential for this technique.

Here we demonstrate full control over an electrostatically defined quantum dot and its tunnel barriers, which allows for recording a time trace of single-electron tunneling on and off the dot, the so-called random telegraph noise enabling us to extract the full counting statistics. It allows us to study all possible correlations and cumulants of charge transfer~\cite{Bagrets2003} as well as the properties reflecting profound aspects of the underlying quantum dynamics and physical mechanisms~\cite{Levitov1996}. The measurement of such a signal requires high sensitivity to single charging events, sufficient signal-to-noise ratio and a high-bandwidth detector with low back-action on the quantum system. We use a capacitively coupled quantum dot (henceforth called the detector) to read out the charge state of the quantum dot (empty or occupied with a single electron). While the full counting statistics of transport through an electrostatically defined quantum dot in GaAs has been demonstrated before~\cite{Gustavsson2006, GUSTAVSSON2009,Flindt2009}, here we present a measurement of the full counting statistics in a gate-defined bilayer graphene quantum dot. We show experimentally the suppression of the second and third cumulants of the distribution of current fluctuations when the quantum dot is symmetrically coupled to the leads. Furthermore, we characterize the spin excited state of the first electron in the quantum dot. The measurements demonstrate the high level of understanding, and control over our graphene devices and pave the way for future experiments in bilayer graphene such as the measurement of degeneracy of charge states via the tunneling rates~\cite{Hofmann2016degeneracy} or entropy~\cite{Hartman2018}. 

\section{Telegraph Signal}

\begin{figure}
	\includegraphics{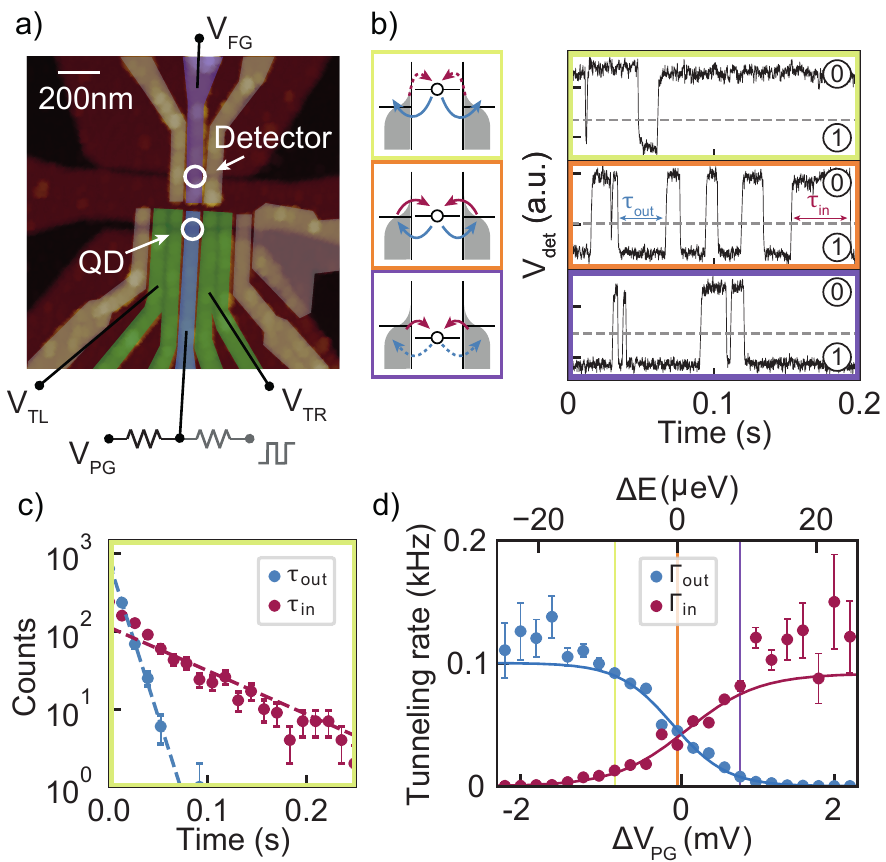}
	\caption{\textbf{(a)} False-color micrograph of the device. The quantum dot (QD) is defined between two tunnel barriers (TL and TR), and its chemical potential is tuned by the plunger gate voltage $V_\mathrm{PG}$. A second capacitively coupled dot formed underneath the finger gate (FG) is used as a charge detector. \textbf{(b)} Time traces of the voltage drop in the detector corresponding to charge fluctuations between an empty dot and one electron in the dot for three different $V_\mathrm{PG}$s and corresponding level schematics. Top: The chemical potential of the quantum dot is above the chemical potential of source and drain. Middle: The chemical potentials of the dot, source and drain are aligned. Bottom: The chemical potential of the quantum dot is below the chemical potential of source and drain. \textbf{(c)} Probability density of the times $\tau_{\mathrm{in}}$ and $\tau_{\mathrm{out}}$ obtained from the time trace shown in the top panel of (b) with a time bin size of $\SI{0.013}{s}$. \textbf{(d)} Evolution of the tunneling rates versus plunger gate detuning from the center of the resonance $\Delta V_\mathrm{PG}$. Fitting a Fermi-Dirac distribution yields an electron temperature of $\SI{52\pm 3}{mK}$.}
	\label{Fig1}
\end{figure}

As shown in Fig.~\ref{Fig1}(a), we define a quantum dot (QD), referred to as the signal dot, between two tunnel barriers (TL and TR) and use a plunger gate (PG) to tune the quantum dot to contain only the first electron. A second quantum dot in the Coulomb blockade regime, referred to as the detector, forms below a single finger gate (FG) in the neighboring channel and is employed as a charge detector.
The same device has been used for the spin-relaxation time measurements presented in Ref.~\onlinecite{GachterGarreis2022} and its geometry is described in detail in the supplemental material S~1.~\cite{supp}.
The detector is biased with a constant current of $\SI{10}{pA}$, the voltage signal $V_\mathrm{det}$ is measured with a detector bandwidth of about $\SI{1}{kHz}$ and sampled with a rate of $\SI{10}{kHz}$. The sample is mounted in a dilution refrigerator with a nominal base temperature of $\SI{10}{mK}$. Unless stated otherwise, all data presented in this manuscript are taken at a perpendicular magnetic field $B_{\perp} = \SI{3.1}{T}$. Here, we expect the single electron ground state to be spin and valley polarized. Due to the finite bandwidth of the detector, a finite magnetic field is necessary to achieve low enough tunneling rates for time resolved charge detection.

A charge carrier tunneling in and out of the signal dot acts capacitively on the detector, hence shifting its conductance resonances~\cite{Kurzmann2019Chargdet}. If an operating point on the rising (falling) edge of a detector resonance is chosen, a step up (down) in the time-dependent detection signal is observed, whenever an electron tunnels off (on) the signal dot. Examples of such time traces are shown in Fig.~\ref{Fig1}(b). In any of these time traces, we can determine the waiting times $\tau_{\mathrm{in}}$ ($\tau_{\mathrm{out}}$) during which the dot is empty (occupied) between two consecutive tunneling events, i.e., the signal is above (below) a predefined threshold (black dashed line). In this device the two levels are well separated and we achieve a signal to noise ratio above 6 (see supplemental material S~2.~\cite{supp}).

For statistically independent tunneling events, the waiting times are exponentially distributed with $p_{\mathrm{in(out)}}(t)\, \text{d}t = \Gamma_{\mathrm{in(out)}} \exp(-\Gamma_{\mathrm{in(out)}} t) \, \text{d}t  $, where $p_{\mathrm{in(out)}}$ is the probability density that an electron enters (leaves) the QD at time $t$ after a complementary event, and $\Gamma_{\mathrm{in(out)}}=1/\langle \tau_{\mathrm{in(out)}} \rangle$ is the tunneling rate of the electron hopping on (off) the quantum dot \cite{Schleser2004}. Here, $\langle \tau_{\mathrm{in(out)}} \rangle$ denotes the statistical mean of the set of experimentally determined waiting times.

Figure \ref{Fig1}(c) shows the measured waiting time distributions for the tunneling-in (-out) events for the time trace, a part of which is shown in the top panel of Fig.~\ref{Fig1}(b). In this measurement, zero source--drain bias voltage was applied to the signal dot, and therefore, the thermodynamic equilibrium state of the quantum dot was probed. The distribution was obtained by binning the set of waiting times $\tau_{\mathrm{in(out)}}$ determined from the time trace. It is evident that the experimental data is well described by the exponential distribution. Since the measurement set-up and detection circuit have a finite bandwidth, all tunnel rates presented in this manuscript are corrected using the finite-bandwidth correction introduced by Naaman and Aumentado \cite{Naaman2006}. In our case, the bandwidth is limited by the detector resistance and cable and filter capacitances. Since the detector resistance depends significantly on the occupation of the dot, the detection bandwidth $\Gamma_{\mathrm{det,up(down)}}$ is also different for tunneling-in and -out (for details see supplemental material S~3.~\cite{supp}). The finite bandwidth correction relates the true tunnel rates $\Gamma_\mathrm{in(out)}$ with the measured rates $\Gamma_{\mathrm{in(out)}}^\ast$ according to 
\begin{align}
\scriptstyle
    \Gamma_{\mathrm{in(out)}} = \frac{\Gamma^\ast_{\mathrm{in(out)}}\Gamma_{\mathrm{det,down}}\Gamma_{\mathrm{det,up}}\pm \Gamma^\ast_{\mathrm{in}}\Gamma^\ast_{\mathrm{out}}\left(\Gamma_{\mathrm{det,up}}-\Gamma_{\mathrm{det,down}}\right)}{\Gamma_{\mathrm{det,down}}\Gamma_{\mathrm{det,up}}-\Gamma^\ast_{\mathrm{in}}\Gamma_{\mathrm{det,up}}-\Gamma^\ast_{\mathrm{out}}\Gamma_{\mathrm{det,down}}}.
\end{align}

For a first characterization of the quantum dot and its tunneling rates, we change the voltage applied on the plunger gate ($V_{\mathrm{PG}}$) and thereby tune the occupation probability of the dot. Figure \ref{Fig1}(b) shows three exemplary time traces. In the top panel the dot resides in the unoccupied state most of the time, in the middle panel the electrons tunnel in and out with similar rates, and in the bottom panel the quantum dot is occupied with one electron most of the time. The schematics depict the corresponding level schemes with arrows indicating the possible tunneling paths. Dashed lines indicate tunneling processes enabled by thermal activation. Here, we only track the charge occupation of the quantum dot, but not the direction of the electron movements.
%because tunnelling processes only occur due to the thermal broadening of the lead potentials. 
Since tunneling to (from) both leads is possible, the measured tunneling rates represent the sum of the tunneling rates between the dot and the left, and the right lead.  
%from (into) which lead the electron enters (leaves) the dot.

Collecting time traces of length $T = \SI{90}{s}$, which is much longer than the waiting times, for several plunger gate voltages, we can map the temperature broadened Fermi-Dirac distribution of the leads. To this end, Fig.~\ref{Fig1}(d) shows the evolution of the tunneling rates $\Gamma_{\mathrm{in(out)}}$ versus plunger gate detuning. While $\Gamma_{\mathrm{in}}$ is seen to increase with increasing plunger gate voltage, $\Gamma_{\mathrm{out}}$ decreases. The qualitative reason of this behavior is the following: If the electrochemical potential of the quantum dot, $\mu_{\mathrm{QD}}$, is energetically above the source and drain electrochemical potentials $\mu_{\mathrm{S/D}}$ at the most negative plunger gate voltages, an electron can tunnel out very quickly due to many available unoccupied states in the leads. At the same time, $\Gamma_{\mathrm{in}}$ is much smaller than $\Gamma_{\mathrm{out}}$ due to the small number of occupied states in the leads above $\mu_{\mathrm{S/D}}$. In analogy, if $\mu_{\mathrm{QD}}$ is resonant with $\mu_{\mathrm{S/D}}$, the tunneling rates $\Gamma_{\mathrm{in}}$ and $\Gamma_{\mathrm{out}}$ are equal, given that the one-electron state in the dot is non-degenerate. When $\mu_{\mathrm{QD}}<\mu_{\mathrm{S/D}}$ (most positive plunger gate voltage), $\Gamma_{\mathrm{in}}>\Gamma_{\mathrm{out}}$.

The gate voltage axis can be converted into an energy axis $\Delta E = e\alpha\Delta V_{\mathrm{PG}}$  using the lever arm $\alpha = 0.113$ determined from finite bias measurements similar to the one shown in Fig.~\ref{Fig4}(a). Note that we use $e>0$ and therefore $\Delta E = \mu -E$. Fitting $\Gamma_{\mathrm{in}}$ and $\Gamma_{\mathrm{out}}$ with temperature broadened Fermi-Dirac distributions $\Gamma_\mathrm{out}^{(0)}f(-\Delta E/k_\mathrm{B}T_\mathrm{e})$ and $\Gamma_{\mathrm{in}}^{(0)}\left[1-f(-\Delta E/k_\mathrm{B}T_\mathrm{e})\right]$ yields an electron temperature $\SI{52 \pm 3}{mK}$, and the tunneling-in and -out rates $\Gamma_{\mathrm{in}}^{(0)}=\SI{0.092\pm0.015}{kHz}$ and $\Gamma_{\mathrm{out}}^{(0)}=\SI{0.10\pm0.01}{kHz}$. We compare the electron temperature obtained from the tunneling rates with the temperature broadened width of the detection step in dc measurements and find that they are in good agreement. We therefore conclude that the detector has a negligible back action on the quantum dot.

\section{Full Counting Statistics}

\begin{figure}
	\includegraphics{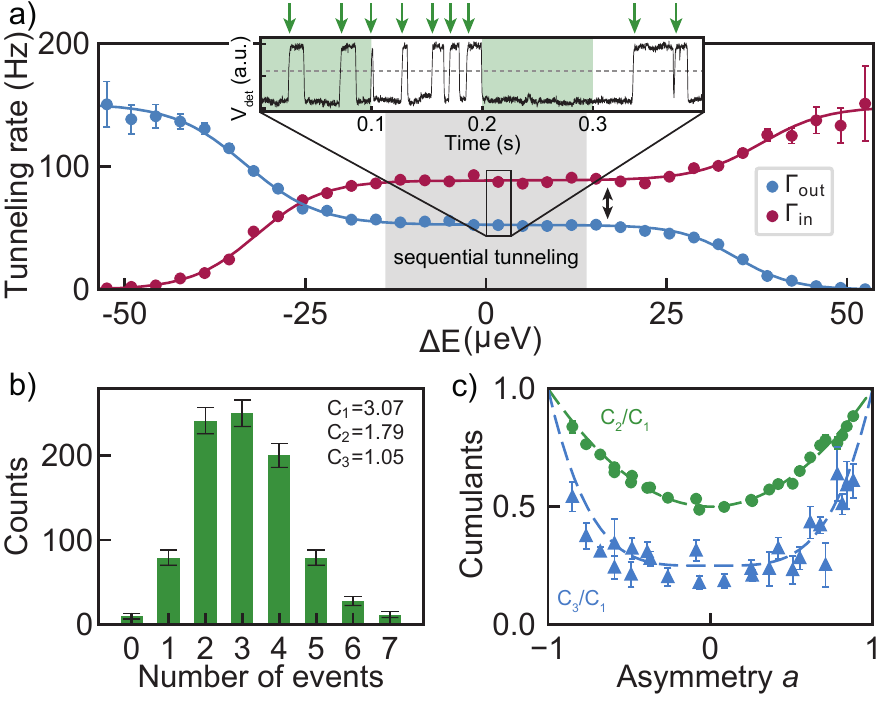}
	\caption{\textbf{(a)} Evolution of the tunneling rates versus energy detuning $\Delta E$ with a finite-bias applied to the dot. The center plateau corresponds to the finite bias window. Inset: Time trace corresponding to an energy detuning around zero. The time trace is divided into sub traces of length $t_0$ and the number of events $n$, i.e. steps up (marked with arrows) per time period is counted. \textbf{(b)} Statistical distribution of the number $n$ of electrons leaving the quantum dot during a given time period $t_0$. We extract the first three cumulants of this distribution. \textbf{(c)} Second and third normalized cumulants of the distribution of $n$ as a function of the asymmetry of the tunneling rates. To improve the statistics, each point at a certain asymmetry is the mean of all time traces obtained within the bias window for given tunnel barrier voltages.}
	\label{Fig2}
\end{figure}

In order to investigate the full counting statistics of the current through the quantum dot, one needs to be able to distinguish electrons tunneling into (out of) the left, or the right lead. This can be achieved by tuning the sample into a non-equilibrium regime, where a finite bias voltage $V_{\mathrm{SD}}$ much larger than the thermal broadening is applied to the signal dot. In this configuration, detecting an electron tunneling out of the dot can be directly associated with a contribution to the current.
We therefore apply a source--drain voltage $V_{\mathrm{SD}}=\SI{75}{\micro V}$ to the signal dot. Figure~\ref{Fig2}(a) shows the resulting tunneling-in and -out rates versus the energy detuning $\Delta E$ of $\mu_{\mathrm{QD}}$ to the midpoint between the source and drain electrochemical potentials. The tunneling rates follow $\Gamma_\mathrm{in} = \Gamma_\mathrm{L}f((\mu_\mathrm{D}-E)/k_\mathrm{B}T_\mathrm{e}) + \Gamma_\mathrm{R}f((\mu_\mathrm{S}-E)/k_\mathrm{B}T_\mathrm{e})$ and $\Gamma_\mathrm{out} = \Gamma_\mathrm{L}\left[1-f((\mu_\mathrm{D}-E)/k_\mathrm{B}T_\mathrm{e})\right] + \Gamma_\mathrm{R}\left[1-f((\mu_\mathrm{S}-E)/k_\mathrm{B}T_\mathrm{e})\right]$, where $\mu_\mathrm{S(D)}$ is the chemical potential of source (drain) and $\Gamma_\mathrm{L(R)}$ the tunneling rate across the left (right) barrier. The shape of the double Fermi distribution reflects the bias window of $\SI{75}{\micro V}$, where the central plateau corresponds to $\mu_{\mathrm{QD}}$ being energetically in-between the source and drain electrochemical potentials. Each step left and right of the center corresponds to $\mu_{\mathrm{QD}}$ being aligned with either the electrochemical potential of source or drain. %Due to electrostatic cross-talk between the gates, the plunger gate influences the tunnel barriers and hence the tunneling rates, which leads to a small slope of the plateau.  
In the sequential tunneling regime, $\Gamma_{\mathrm{in(out)}}$ is purely dominated by the right (left) tunnel barrier, hence the difference of the two tunneling rates between the Fermi steps, marked with a black double-arrow, is related to the asymmetry $a = (\Gamma_{\mathrm{R}}-\Gamma_{\mathrm{L}})/(\Gamma_{\mathrm{R}}+\Gamma_{\mathrm{L}})$ between the two barriers. Here $\Gamma_\mathrm{R (L)}$ denotes the mean of the tunneling rates in the sequential tunneling regime labeled in Fig.~\ref{Fig2}.

To investigate the statistical properties of sequential electron transport through the quantum dot, we analyze time traces for energies $\Delta E$ well within the bias window. As shown in the inset of Fig.~\ref{Fig2}(a), we divide the time trace into sub-traces of length $t_0$ and count the number of events $n$, i.e., steps up (marked with arrows) per time period. Evaluating $T/t_0$ sub-traces yields the histogram shown in Fig.~\ref{Fig2}(b), i.e., the full counting statistics of the current, where we plot $n$ on the horizontal axis and the counts of sub-traces with $n$ tunneling events on the vertical axis. Using this distribution, we determine the first three cumulants: $C_1$, the mean $\langle n \rangle$, and $C_i = \langle (n-\langle n  \rangle)^i \rangle$ for $i = 2,3$. These cumulants are related to the mean current flowing through the dot (1st cumulant), and the shot noise $S_I$ (2nd and higher cumulants) which specifies the statistical current fluctuations in the quantum dot~\cite{GUSTAVSSON2009}.

We repeat the measurement of the full counting statistics of the tunneling current for different tunneling asymmetries $a$. At each value of $a$, we measure the full counting statistics several times. From each of them the statistical cumulants are determined for a specific value of $a$. Then their mean values and the uncertainty of the mean are computed. Figure~\ref{Fig2}(c) shows the resulting Fano factors $\frac{C_2}{C_1}$, as well as the ratio $\frac{C_3}{C_1}$ versus the asymmetry~$a$. Compared to a purely Poissonian process, the cumulants are expected to be reduced for a Coulomb-blockaded system, since an electron can only enter the dot if the previous one has left, which leads to correlations in electron tunneling. It is theoretically expected that the normalized cumulants follow $C_2/C_1 = 0.5(1+a^2)$ and $C_3/C_1 = 0.25(1+3a^4)$~\cite{Bagrets2003}. These theoretical expressions [dashed lines in Fig.~\ref{Fig2}(c)] describe the experimentally obtained data very well, with no fitting parameters involved. Note that most data points lie slightly below the predicted curve: due to the finite measurement bandwidth, events very close to the border between two time periods $t_0$ are associated with the wrong time bin.

\begin{figure}
	\includegraphics{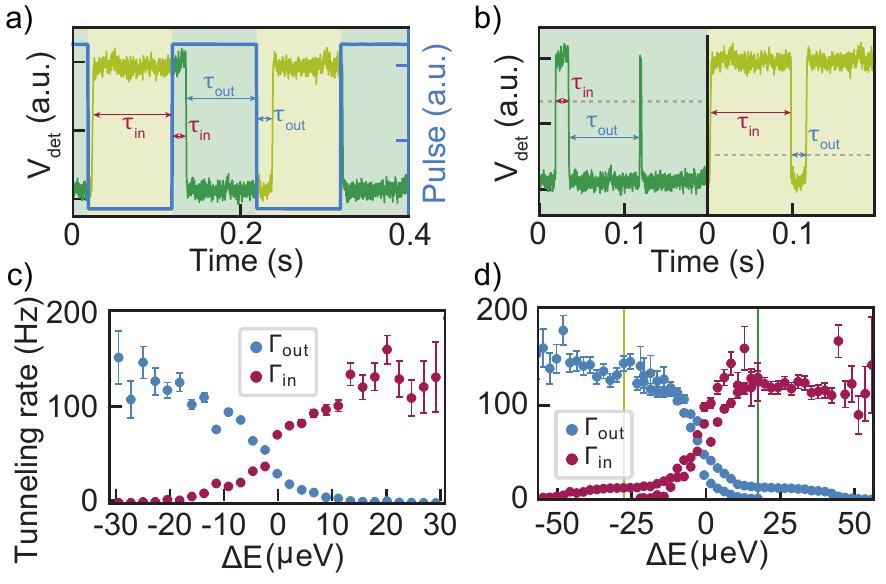}
	\caption{\textbf{(a)} Time trace for a two-level pulse applied to the plunger gate. The signal is color coded by the parts where the electrochemical potential of the quantum dot is above (olive) and below (green) the electrochemical potential of source and drain. The pulse sequence is plotted in blue. \textbf{(b)} Same time trace as shown in (a), but separated into the two regimes of total $\Delta E$ to be analyzed individually. \textbf{(c)} Tunneling rates versus energy detuning for an unpulsed measurement. \textbf{(d)} Tunneling rates for the same gate configuration as in (c), but with an additional square pulse applied to the plunger gate. The resolution for energies where the Fermi-Dirac distribution is at the higher plateau is increased considerably. The green and olive lines mark the pulsed energy configuration of the trace shown in (a) and (b).}
	\label{Fig3}
\end{figure}

\section{Excited State Spectroscopy}

In Fig.~\ref{Fig3}(c), for detunings above $\SI{15}{\micro eV}$ the uncertainties of the tunneling-in rates drastically increase. Here, the tunneling-in and -out rates are very asymmetric, hence the number of events for filling a mostly filled level becomes very small. The investigation of excited states needs a precise measurement with high statistics of the tunneling rates over a large detuning range, hence also for very asymmetric tunneling-in and -out rates. Therefore, we add a two-level pulse to the plunger gate voltage with a frequency $\nu_{\mathrm{pulse}} = \SI{2}{Hz}$. At this frequency, the waiting time for tunneling-in at $\Delta E>0$ is much smaller than half the duty cycle, whereas the waiting time for tunneling out is much larger. A similar argument applies to the case where $\Delta E<0$. The pulse amplitude of $\SI{22.6}{\micro eV}$ is much larger than the step sizes in dc plunger gate voltage used for the sweep in Fig.~\ref{Fig3}(d). In case of $\mu_\mathrm{QD}$ below the electrochemical potential of source and drain, this shortens the long waiting time after a tunneling-in event by pulsing the level so much up in energy that it is quickly filled. The level is then pulsed down again to the desired energy. This induces two tunnel events for each pulse at two different total plunger gate voltages and improves the number of events detected within a given measurement time \cite{Elzerman2004_exc}. This method measures fast tunneling rates more efficiently, while very slow tunneling rates can not be measured accurately.
%%%

In our experiment, the square pulse is combined with the dc offset $V_{\mathrm{PG}}$ via two resistors at room temperature as depicted in Fig.~\ref{Fig1}(a). Figure~\ref{Fig3}(a) shows an exemplary resulting time trace (olive and green) and the applied pulse sequence (blue) switching between two different total plunger gate voltages. In the green gate configuration the dot is statistically more likely to be occupied, vice versa in the olive gate configuration it is likely to be empty. Immediately after the pulse step up (down), the dot occupation stays constant for a finite amount of time until a tunneling-out (-in) event occurs. Similar to the non-pulsed case, this wait time can be converted into a tunneling rate. Based on the pulse frequency, the trace is rearranged and split into two time traces corresponding to two different absolute gate voltages, and are evaluated separately [Fig.~\ref{Fig3}(b)]. The advantage of this measurement technique is demonstrated by comparing Figs.~\ref{Fig3}(c) and (d). Since the tunneling events’ statistic is improved, the measured $\Gamma_{\mathrm{in(out)}}$ has a lower uncertainty. Furthermore, reliable tunneling rates can be measured for a larger range of $\Delta E$, because the events are no longer limited by thermal broadening of the leads, but by the pulse amplitude. Note that the lower tunneling rate in the pulsing-induced tunneling regime [e.g., $\Gamma_{\mathrm{out}}$ around the green line in Fig.~\ref{Fig3}(d)] does not go to zero entirely, but reaches a constant value of $\Gamma = 2 \nu_{\mathrm{pulse}}$. Additionally, for $|\Delta E| < \SI{20}{\micro eV}$, each tunneling rate is evaluated from two different pulsed traces measured at two different absolute plunger gate voltages. Charge fluctuations in the sample, resulting in small shifts in $\Delta E$, yield a small offset between the two data sets. These two effects have no impact on the following discussions, since we are mostly interested in the higher tunneling rates.

\begin{figure}
	\includegraphics{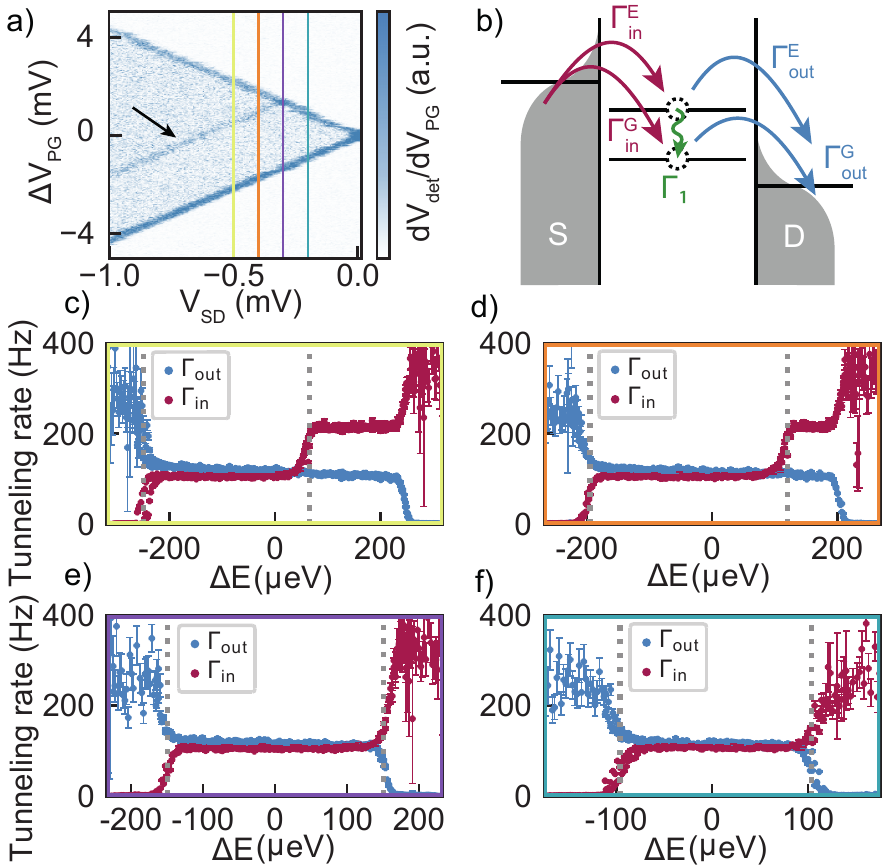}
	\caption{\textbf{(a)} Measured $\text{d}V_{\mathrm{det}}/\text{d}V_{\mathrm{PG}}$ at finite bias for the first charge carrier transition. The spin excited state is marked with an arrow. \textbf{(b)} Schematic energy diagram of the quantum dot including the first spin excited state. Possible tunneling and relaxation paths are labeled. \textbf{(c)-(f)} Tunneling rates versus energy detuning for different source drain biases as indicated in (a). As soon as the spin excited state enters the bias window, the additional tunneling path yields an increased tunneling-in rate.}
	\label{Fig4}
\end{figure}

The pulsing technique also allows for $\text{d}V_{\mathrm{det}}/\text{d}V_{\mathrm{PG}}$ measurements with a pulse frequency of $\SI{205}{Hz}$, which is in the same order of magnitude as the tunneling rates. The pulse amplitude is $\SI{290}{\micro V}$. In this domain, the differential change of the quantum dot occupation is measured, if $V_{\mathrm{det}}$ is demodulated at the pulse frequency. Figure~\ref{Fig4}(a) shows a finite bias measurement of the first electron in the quantum dot at a perpendicular magnetic field $B_{\perp} = \SI{2.1}{T}$. For large enough bias, the first spin excited state appears in the bias window (at the energy corresponding to the sum of zero-field spin--orbit splitting and Zeeman splitting~\cite{GachterGarreis2022}), which increases the overall current through the dot, resulting in a peak in the $\text{d}V_{\mathrm{det}}/\text{d}V_{\mathrm{PG}}$ signal. As depicted in the schematics in Fig.~\ref{Fig4}(b), in the $V_{\mathrm{SD}}-\Delta V_{\mathrm{PG}}$ region above this line, the excited state offers an additional channel to tunnel onto the quantum dot and the tunneling-in rate is expected to increase. However, due to Coulomb blockade, only one electron can enter the dot at a time, hence the tunneling rate $\Gamma_{\mathrm{out}}$ stays constant if the tunneling out rates of the ground and excited states are equal. Figures~\ref{Fig4}(c)-(f) show the tunneling rates extracted from time traces collected at the different line cuts marked in Fig.~\ref{Fig4}(a). 

In the following, we discuss the tunneling rates of the ground ($\Gamma^{\mathrm{G}}$) and the excited state ($\Gamma^{\mathrm{E}}$) more quantitatively. Suppose only the ground state is within the bias window [the region enclosed by two dotted lines in Fig.~\ref{Fig4}(c-f)], then only one state participates in tunneling. The tunneling rates are given by $\Gamma_{\mathrm{in}} = \Gamma^{\mathrm{G}}_{\mathrm{in}}$ and $\Gamma_{\mathrm{out}} = \Gamma^{\mathrm{G}}_{\mathrm{out}}$. Once the excited state enters the bias window, the tunneling-in rate increases to $\Gamma_{\mathrm{in}} = \Gamma^{\mathrm{G}}_{\mathrm{in}} + \Gamma^{\mathrm{E}}_{\mathrm{in}}$. To calculate the tunneling-out rate, the probabilities for three different alternative processes need to be considered: the electron entered in the ground state and leaves the quantum dot with rate $\Gamma^{\mathrm{G}}_{\mathrm{out}}$; the electron entered in the excited state and leaves the quantum dot with $\Gamma^{\mathrm{E}}_{\mathrm{out}}$; or, the electron entered the excited state and relaxed into the ground state with $\Gamma_{1}$ and then leaves the quantum dot with $\Gamma^{\mathrm{G}}_{\mathrm{out}}$. This yields an overall tunneling out rate of: 
\begin{align}
  \Gamma_{\mathrm{out}} = \frac{1}{\langle\tau_\mathrm{out}\rangle} = \frac{\Gamma^{\mathrm{G}}_{\mathrm{out}}(\Gamma^{\mathrm{G}}_{\mathrm{in}}+\Gamma^{\mathrm{E}}_{\mathrm{in}})(\Gamma^{\mathrm{E}}_{\mathrm{out}}+\Gamma_1)}{\Gamma^{\mathrm{E}}_{\mathrm{in}}(\Gamma^{\mathrm{G}}_{\mathrm{out}}+\Gamma_1)+\Gamma^{\mathrm{G}}_{\mathrm{in}}(\Gamma^{\mathrm{E}}_{\mathrm{out}}+\Gamma_{1})}. 
\end{align}
Hence, as soon as tunneling to the excited state is possible, a step in tunneling-out rate of 
\begin{align}
    \Delta \Gamma_{\mathrm{out}} = \Gamma^{\mathrm{G}}_{\mathrm{out}} - \Gamma_{\mathrm{out}} = \frac{\Gamma^{\mathrm{G}}_{\mathrm{out}}\Gamma^{\mathrm{E}}_{\mathrm{in}}(\Gamma^{\mathrm{G}}_{\mathrm{out}}-\Gamma^{\mathrm{E}}_{\mathrm{out}})}{\Gamma^{\mathrm{E}}_{\mathrm{in}}(\Gamma^{\mathrm{G}}_{\mathrm{out}}+\Gamma_1)+\Gamma^{\mathrm{G}}_{\mathrm{in}}(\Gamma^{\mathrm{E}}_{\mathrm{out}}+\Gamma_{1})}
    \label{deltagammaout}
\end{align}
is expected. The measurements show $\Delta \Gamma_{\mathrm{out}} = 0$, hence assuming $\Gamma^{\mathrm{G}}_{\mathrm{out}} \Gamma^{\mathrm{E}}_{\mathrm{in}} \neq 0$ yields $\Gamma^{\mathrm{G}}_{\mathrm{out}} = \Gamma^{\mathrm{E}}_{\mathrm{out}}$. This means that in our experiment the spin-up and spin-down states of the one-electron quantum dot share identical tunneling-out rates within our measurement precision. In comparison, similar experiments in GaAs show different tunneling in rates for different excited states~\cite{Gustavsson2006timeresolved} as well as Pauli spin blockade in double dots~\cite{Hofmann2017}. Here, we assume that the relaxation rate in Eq.~\eqref{deltagammaout} is not much bigger than the tunneling rates. We validate this assumption by direct measurements of the relaxation rate $\Gamma_1 = \SI{120}{Hz}$ using the Elzerman spin readout technique~\cite{Elzerman2004}, which has been presented in Ref.~\onlinecite{GachterGarreis2022} for this device and gate configuration.

\section{Conclusion}

In conclusion, we have demonstrated time-resolved measurements of an electron occupying and leaving a quantum dot in bilayer graphene. The high sample quality and sensitivity of the measurement technique allow us to detect random telegraph noise, which we analyse by looking at the the full counting statistics. We also used a square pulse to improve the statistics in the non-equilibrium gate configuration, which enables the investigation of the first spin excited state and is crucial for spin relaxation measurements as demonstrated in Ref.~\onlinecite{GachterGarreis2022}.\\

\section*{Acknowledgement}
We are grateful for the technical support by Peter M\"arki, Thomas B\"ahler. We acknowledge financial support by the European Graphene Flagship, the ERC Synergy Grant Quantropy, the European Union’s Horizon 2020 research and innovation programme under grant agreement number 862660/QUANTUM E LEAPS and NCCR QSIT (Swiss National Science Foundation, grant number 51NF40-185902) and under the Marie Sklodowska-Curie Grant Agreement Number 766025. K.W. and T.T. acknowledge support from the Elemental Strategy Initiative conducted by the MEXT, Japan, Grant Number JPMXP0112101001, JSPS KAKENHI Grant Number JP20H00354 and the CREST(JPMJCR15F3), J.S.T.

%\bibliography{library}


\title{Counting Statistics of Single Electron Transport in Bilayer Graphene Quantum Dots:\\ Supplemental Material}

\date{\today}
	
\maketitle
	
\newpage

\setcounter{section}{0} 

\renewcommand\thesection{S~\arabic{section}} 
\setcounter{figure}{0} 
\renewcommand\thefigure{S\arabic{figure}} 

\section{Sample Geometry}

\begin{figure}
	\includegraphics{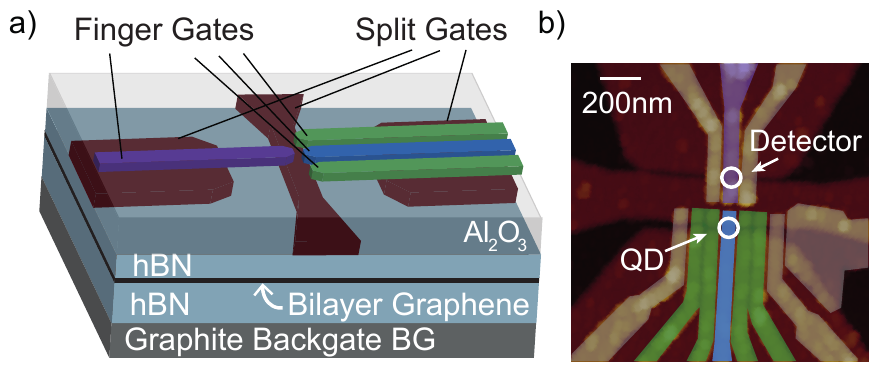}
	\caption{\textbf{(a)} Schematic representation of the stack. From bottom to top it is built up with a graphite back gate (grey), a bottom hBN (light blue), a bilayer graphene flake capped with a top hBN. Separated vertically by an aluminium oxide layer, the split gates (brown) and finger gates are used to electrostatically define a quantum dot with tunnel barriers (green) below the plunger gate (blue) and a detector dot (purple).  (b) False color atomic force microscope picture of the sample, with split gates (brown) and barrier (green) defined quantum dot (blue) and detector (purple). The other finger gates (grey) are grounded.}
	\label{Fig_S1}
\end{figure}

The fabrication of the heterostructure shown schematically in Fig.~\ref{Fig_S1}(a) follows the general procedure described in previous publications \cite{Overweg2018, Eich2018, Banszerus2018}. From bottom to top it is built up with a graphite back gate, a bottom hBN ($\SI{31}{nm}$), a bilayer graphene flake capped with a top hBN ($\SI{20}{nm}$). The split gates ($\SI{5}{nm}$ Cr and $\SI{20}{nm}$ Au) are designed to form two channels of a nominal width of $\SI{100}{nm}$ with a separation gate of \SI{150}{nm} width in between. The finger gates ($\SI{5}{nm}$ Cr and $\SI{20}{nm}$ Au) have a width of $\SI{20}{nm}$ and a center to center distance of $\SI{85}{nm}$. The aluminium oxide layer (Al$_2$O$_3$) separating the two metallic gate layers has a thickness of $\SI{30}{nm}$. Figure~\ref{Fig_S1}(b) shows a false color atomic force microscope image of the two layers of metal gates fabricated on top of the heterostructure. The split gates [brown in Fig.~\ref{Fig_S1}(a),(b)] are used to form two conducting channels (black in Fig.~\ref{Fig_S1}(b)) \cite{Overweg2018}. For the measurements discussed in this paper we use a single finger gate (purple) to define a p-n-junction defined quantum dot~\cite{Eich2018,Banszerus2018} in the upper channel which we utilize as a charge detector~\cite{Kurzmann2019Chargdet}. In the second channel we define a quantum dot (QD) below one finger gate (blue) and use the neighboring gates (green) to tune the tunnel coupling of the quantum dot to the leads~\cite{Tong2021, Banszerus2020}. All other gates are grounded.

\section{Characterization of the Signal to noise ratio}
\begin{figure}
	\includegraphics{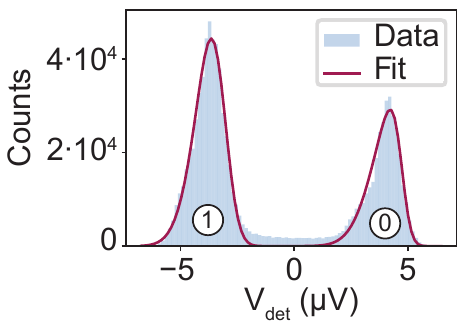}
	\caption{Histogram of detector voltage values during a time trace of $\SI{90}{s}$. The histogram is fitted with two skewed Gaussians with $\gamma_\mathrm{0}= -3.21$, $\mu_\mathrm{0} = \SI{4.72}{\micro V}$, $\sigma_\mathrm{0} = \SI{-1.08}{\micro V}$ and $\gamma_\mathrm{1}= 1.80$, $\mu_\mathrm{1} = \SI{-3.11}{\micro V} $, $\sigma_\mathrm{1} = \SI{-0.98}{\micro V}$, where $0,1$ denote the electron occupation of the dot.}
	\label{Fig_S2}
\end{figure}

Analyzing an exemplary time trace of \SI{90}{s} we extract the histogram of the peak values of the detector voltage $V_\mathrm{det}$ shown in Fig~\ref{Fig_S2}. We identify two well-separated peaks indicating that $V_\mathrm{det}$ has two favorable values corresponding to the first electron being or not being in the QD. The two voltage levels are separated with a signal to noise ratio $\text{SNR}=\left|\mu_1-\mu_2\right|/{\sqrt{\sigma_1^2+\sigma_2^2}}=6.3$ as obtained from fitting two skewed Gaussians $A \exp\left[-(x-\mu)^2/2\sigma^2\right][1-\mathrm{erf}(\gamma(x-\mu)/\sqrt{2}\sigma)]$.

\section{Evaluation of the Detection Bandwidth}

\begin{figure}
	\includegraphics{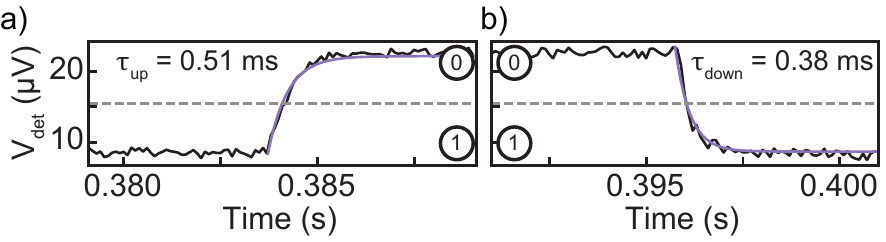}
	\caption{\textbf{(a)} Response time of the detector for an electron leaving and \textbf{(b)} an electron entering the quantum dot. The signal is fitted with $\exp(-t/\tau_\mathrm{up(down)})$, where $\Gamma_\mathrm{det, up(down)} = 1/\langle \tau_\mathrm{up(down)} \rangle$ describes the detection bandwidth for the respective event.}
	\label{Fig_S3}
\end{figure}

When an electron enters the quantum dot, the detector potential and thus its resistance changes abruptly. However, due to parasitic resistances and capacitances, $V_\mathrm{det}$ reacts with a characteristic response time $\tau$. In Fig.~\ref{Fig_S3} this exponential course is visible for an electron leaving (Fig.~\ref{Fig_S3}(a)) and entering the signal dot (Fig.~\ref{Fig_S3}(b)). As illustrated, the response time for an up and a down transition differ, which can be explained as follows: The resistance of the detector is a dominant factor in the RC response time of an detection event. Hence, a big resistance change varied the detection bandwidth significantly. The response time $\tau$ originates from an RC time constant with a constant parasitic capacitance $C = \Delta \tau /\Delta R$. For the traces shown in Fig.~\ref{Fig_S3} we get $C = 0.19 \pm 0.08\,\text{nF}$, which is consistent with the line capacity $C_\mathrm{line} = \SI{0.1}{nF}$ of the source and drain leads.

For the bandwidth correction of the tunneling rates presented in the main text, we use one $\Gamma_\mathrm{det, up(down)}$ for one data set of varying plunger gates. We choose several time traces of plunger gate voltages where the tunneling rates are roughly symmetric to obtain reasonable statistics and fit each transition with an exponential $\exp(-t/\tau_\mathrm{up(down)})$. The respective detection bandwidth for the corresponding data set is then given by $\Gamma_\mathrm{det, up(down)} = 1/\langle \tau_\mathrm{up(down)} \rangle$. The measurements for the fano factor were conducted with similar detector tuning. We therefore correct this data set with one single averaged $\Gamma_\mathrm{det, up} = \SI{0.60 \pm 0.08}{ms}$ and $\Gamma_\mathrm{det, down} = \SI{0.34 \pm 0.08}{ms}$.

%\bibliography{literatur} 

\end{document}